# Devising Malware Characteristics using Transformers

Simra Shahid[1], Tanmay Singh[2], Yash Sharma[3], Kapil Sharma[4]

[1]simrashahid_bt2k16@dtu.ac.in, [2]tanmay_bt2k16@dtu.ac.in, [3]yash4688@gmail.com, [4]kapil@ieee.org

***Abstract***— *In this paper, we present our approach of finding relevant malware behaviour texts from Malware Threat Reports as described by Lim [1]. Our main contribution is the opening attempt of Transfer Learning approaches, and how they generalize for the classification tasks like malware behaviour analysis.*

## I. INTRODUCTION

The digital landscape is unique and constantly changing, creating room for cyber-attacks. Amidst the rise of these security threats and vulnerabilities, it has become crucial to identify it and take the appropriate action. With the advancements in the digital landscape, we need better security tools to combat different kinds of threats. By 2019, there was a 13% rise in pre-installed malware and adware on Android devices, and what's even more shocking, the Macs, which are known for its durable security barriers, had more threats detected than Windows. These attacks play with one's data, money and privacy. This urges the large scale companies and customers to effectively know where they are at alerting, blocking and detecting threats.

Advance Persistent Threat (APT) is a targeted attack in which an intruder gains access to a network to monitor network activity and steal data rather than to cause damage to the network or organization. APT reports contain detailed malware behaviour analysis of their onset and traversal through our system when under attack. Despite the extant repositories of these malware [2], it becomes difficult for security researchers to skim through these huge databases to find useful content. Reading through the massive documents, make it impossible to analyse and quickly act upon the adversary. There is a need to be able to automate this process without having to read through the entire report.

The advent of these malware threat reports engenders the security analysts to make use of natural language processing (NLP) algorithms, to identify, cluster, analyse the pattern of malware [3]. Keeping this in mind, we propose some malware text isolation techniques to detect such sentences, and further develop a relationship between them.

Upon delineating sentences depicting malware [4] behaviour capabilities from the large volume of threat reports will help security analysts to quickly decide evasion strategies from the threat landscape, clustering malware [5] which have similar behaviour, learn about the security vulnerabilities of the enterprise and strengthen the system from similar future attacks.

Managing the extraction of such sentences from large corpora, we can generalize and cluster similar Malwares together, further helping new researches to analyse and reduce the vulnerabilities in the networks. Albeit the applications are endless, not much work has been done in conjunction with applying Natural Language Processing and Malware [6] Analysis as in the following references.

For example, the following sentence does not depict malware behaviour:

*Once decoded, FireEye identified the payload as a poison ivy variant.*

Whereas, the next sentence is describing the course of action upon attack by an intruder:

*The backdoor contained versioning info which attempted to masquerade as a Google Chrome File.*

Our main contributions are:
1) We introduce transformers approaches like ULMFiT, BERT, XLNETs for the malware characteristics classification task.
2) We discuss different sampling approaches to the class imbalance.
3) We make an opening attempt in investigating the effectiveness of transfer learning for the problems in the domain of security.

## II. RELATED WORK

In 2018, SemEval organized a shared task called SecureNLP on semantic analysis for cybersecurity texts 1. Task 1 was a binary classification task of sentences extracted from APT reports which had malware behaviour or not. In this section, we briefly describe the approaches by the competition for the task.

Using Glove embeddings proposed by Pennington [7], Villani [8], outperformed the rest of the competition in Subtask 1 only. With Long Short Term Memory network (LSTM), they generated token representation from the characters. Following that, a binary classifier was trained with Bi-directional Long Short-Term Memory network (BiLSTM).

Flytxt NTNU [9] assembled an ensemble of Conditional Random Field (CRF) and Naive Bayes classifier for SubTask 1. The CRF model used lexical-based and context-based features. If the CRF predicts any "BIO" token labels (SubTask2) for the sentence, the sentence is considered relevant in SubTask 1.

DM-NLP [10] used the predicted output labels from SubTask 2 to get the predictions for SubTask 1. They model this task as a sequence labelling task and used a hybrid approach with BiLSTM-CNNCRF as mentioned in [11].

HCCL [12] performed a very similar approach to team DM-NLP using the same BiLSTM-CNN-CRF architecture. They used relatively simpler Part-Of-Speech (POS) features, instead of the more complicated linguistic features like the former team. They aim to build an end-to-end system without any feature engineering or data preprocessing.

Digital Operatives [13] utilized a passive-aggressive classifier Reference [14], which has comparable cost and performance with the linear Support Vector Machine classifier, for SubTask 1. The features they applied include POS, dependency links, and bigrams.

TeamDL [15] built a convolutional neural network with original glove embeddings. UMBC [16] used a Multilayer Perceptron model for the submission of SubTask 1. Inspired from the tasks, Ravikiran [17] has proposed a multimodal dataset with QR-codes and Malware Text classification.

In this paper, we particularly focus on language modelling approach for the malware behaviour classification. The following section describes our approaches.

## III. OUR METHODOLOGY

In this section, we aim to discuss in detail our approach to solve subtask 1. After analysing the previously designed models, our team worked on some new approaches towards the challenge. The following section discusses the details of the steps used for the construction of these experiments.

### A. Preprocessing

We preprocessed the models by removing punctuations, numbers, and did the following modifications:
- .exe like copy.exe files to [EXE].
- Buffer memory and Stack memory addresses like 0x20000001 are replaced by [ADDRESS].
- Malware TrojanDropper.Win32.Agent.life name replaced by [MALWARE].
- .bat, .doc, .txt file names replaced by [FILE].
- File paths to [PATH].
- IP Addresses to [IP].

We further removed texts which were not in English and had only numbers.

### B. SubTask1: Malware Threat Classification

This section gives an in-depth detail of the transformer models used for the classification task for SubTask1. The models aimed to extract the cybersecurity-related terms from a sentence and then classify them into one of the two classes: malware related or non-malware related. Figure 1 shows the model architecture.

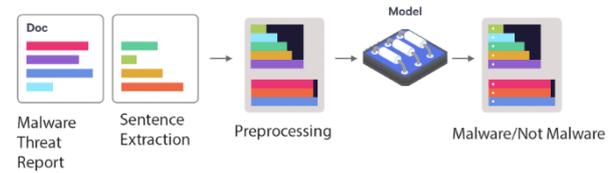

**Fig. 1** Model Architecture

*1) ULMFit:* [18] is short for Universal Language Model Fine-Tuning for Text Classification. The authors' brought out the disadvantages of using traditional word embedding approaches [19] directly with deep neural networks. The random initialization of Out of Vocabulary (OOV) Words which disrupts the pre-trained layers and causes catastrophic forgetting. To overcome that the paper discusses different approaches of gradual freezing and discriminate fine-tuning. The ULMFiT's backbone is divided into the following stages:

1) Language Model pre-training

2) Language Model fine-tuning

3) Classifier Model fine-tuning

It is a universal model as it works with varying document sizes, requires no custom feature engineering nor preprocessing, and uses a single architecture. AWD-LSTM language model is used in the architecture which comprises a conventional LSTM with no added attention. We tried experimenting with the data for the language model, adding domain-specific data. But this addition didn't account for any improvement.

ULMFiT's conventional parameters of fast.ai were used to train the language models. Finally, we find the best hyperparameters by learning rate finder and train the classifier over the task data.

The problem with ULMFiT is the words in a sentence are sequentially processed and still does not capture the true meaning of the context

*2) BERT:* Bidirectional Encoder Representations from Transformers (BERT) [20] was the first language model which is deeply bidirectional, unsupervised language representation, pre-trained using only a plain text corpus. This model takes the entire context and processes it simultaneously, capturing the true context of a word.

Transformer has two mechanisms - an encoder and a decoder. Encoder reads the text input and the Decoder gives a prediction for the intended task. BERT makes use of the Transformer's Encoder Architecture, which has an attention mechanism that learns contextual relations between words in a sentence.

It performs two tasks for language modelling:

1) Masked Language Modelling Bert randomly [MASKS] a word and predicts it using its context from left and right simultaneously. This masked language model (MLM) learns to model relationships between words and sentences.

2) Next Sentence Prediction Model This model takes two sentences as its input S1 and S2, and verifies whether S2 follows S1, capturing the relationship between sentences.

For this challenge, we have used the pre-trained Hugging Face implementation of the BERT-base model and fine-tuned it for our task dataset.

3) *XLNets:* XLNET [21] has a very similar architecture, as of BERT. It uses a different approach of masking and uses Transformer XL model instead of a Transformer model. Instead of masked language modelling, XLNET uses permutation language modelling (PLM). It blends the concept of autoregressive models and bidirectional context modelling. PLM is the idea of capturing a bidirectional context by training an autoregressive model on all possible permutation of words in a sentence. Instead of fixed left-right or right-left modelling, XLNET maximizes expected log-likelihood over all possible permutations of the sequence. In expectation, each position learns to utilize contextual information from all positions thereby capturing bidirectional context. No [MASK] is needed and input data need not be corrupted.

## IV. EVALUATION

This section gives a detailed overview of the dataset introduced in the SubTask1, Semeval Task 8: SecureNLP Challenge. Since the data was highly imbalanced we tried various undersampling and oversampling approaches. This section concludes with details of hyperparameters used, metric chosen for evaluation and the results obtained.

**Table I** Dataset

|  | Documents | Sentence |
|---|---|---|
| Train | 65 | 9,424 |
| Dev | 5 | 1,213 |
| SubTask1 test | 5 | 618 |
| Total | 75 | 11,250 |

### A. Dataset

The total statistics of the dataset are shown in Table I. The training data for this shared task contains 9,424 sentences, the validation data contains 1,213 sentences, and test data has 618 test sentences. Figure 2 shows the huge class imbalance between malware related/non-malware related sentences.

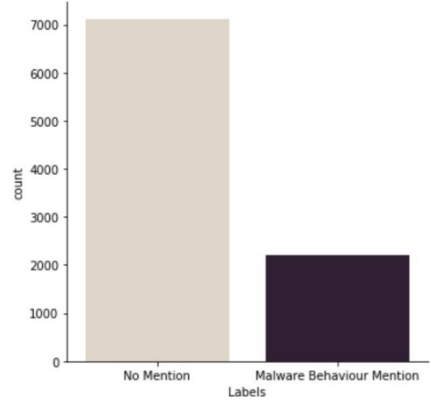

Fig. 2. Dataset Frequency

### B. Metrics

The evaluation metric chosen by the Challenge to evaluate the performance of the malware classification task was f1 score. We have computed the Precision and Recall as well. Precision is the fraction of relevant instances among the total retrieved instances. Recall is the fraction of relevant instances retrieved over the total amount of relevant instances. Precision is computed as:

$$Precision = \frac{TP}{TP + FP}$$

Recall is computed as:

$$Recall = \frac{TP}{TP + FN}$$

F1 score is computed using precision and recall as follows:

$$F1 = 2 \times \frac{Precision \times Recall}{Precision + Recall}$$

The systems were learned from the training validation data and tested on the evaluation data.

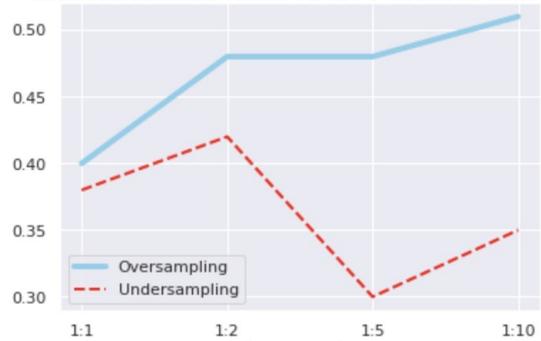

Fig. 3. Sampling Ratio

### C. Sampling

Now that we know how to set up the evaluation scheme and what metrics to choose for classification problems with imbalanced data, we apply some techniques to account for class imbalance. The most straightforward technique is to balance the data by resampling:

1) Down-sampling (Under sampling) the majority class

2) Up-sampling (Over sampling) the minority class

Without resampling the data, one can also make the classifier aware of the imbalanced data by incorporating the weights of the classes into the cost function. Intuitively, we want to give higher weight to minority class and lower weight to the majority class. We have tried class weight techniques for BERT. While training the Language Model for ULMFiT, We observed that fine-tuning the language model on a larger dataset didn't provide much improvement. For more data, we scraped threat reports from MalwareTextDB1.0. We finetuned the language model over 6200 (scraped data) + 11,250 (original data). For scraping, we used the tokenized data and assigned a sentence class 0 when all the BIO token-indexes were O, else we assigned it to class 1. In Table III we show the hyperparameters which yielded good performance models. In Figure 2, we show the different sampling ratios used against the F1 Score predicted for both oversampling and undersampling models using BERT-cased.

Table II Experimental Results on SubTask 1

| Model | Remarks | Epochs | Precision | Recall | F1 score |
|---|---|---|---|---|---|
| BERT | Oversampling 1:10 | 6 | 0.37 | 0.85 | 0.51 |
| BERT | Oversampling 1:2 | 10 | 0.31 | 0.85 | 0.46 |
| BERT | Oversampling 1:2 | 6 | 0.32 | 0.96 | 0.48 |
| BERT | Oversampling 1:2 | 3 | 0.33 | 0.93 | 0.48 |
| **BERT** | **No sampling** | **5** | **0.49** | **0.55** | **0.52** |
| BERT | Undersampling 1:1 | 5 | 0.62 | 0.28 | 0.38 |
| XLNET | Undersampling 1:2 | 5 | 0.29 | 0.89 | 0.44 |
| XLNET | No sampling | 10 | 0.36 | 0.64 | 0.46 |
| XLNET | No sampling | 4 | 0.26 | 0.88 | 0.41 |
| ULMFIT | No sampling, LM with **same** dataset | 20 | 0.74 | 0.25 | 0.38 |
| ULMFIT | No sampling, LM with **same** dataset | 30 | 0.42 | 0.48 | 0.45 |
| ULMFIT | No sampling, LM with **same** dataset | 50 | 0.30 | 0.48 | 0.37 |
| ULMFIT | LM with **larger** dataset | 5-10 | 0.90 | 0.16 | 0.27 |

### D. Result

Table II shows the comparison of F1 scores on test data for SubTask1. The BERT-cased model showed the best performance amongst the different transformers. Figure 3, presents the performance of the BERT-cased model on the training data for different oversampling and undersampling ratios. We observed that oversampling performance was better than undersampling. However, the results without any sampling were the best one. Therefore, the model that we used on the test data was trained on the full training dataset maintaining the given ratio of malware: non-malware tweets. Table III shows the hyperparameters of the best running models.

Table III Hyperparameters of Best Performing Models

| Model | Hyperparameters |
|---|---|
| BERT | epochs=5, batch size=32, learning rate=3e-5 |
| ULMFit | epochs= 30, batch size=32, learning rate = 2e-6 |

### V. CONCLUSION

In this paper, we present a transformer approach targeting SemEval 2018 shared task on Semantic Extraction from CybersecUrity REports using Natural Language Processing (SecureNLP). We were able to produce a model that generates feasible results for estimating the relevance of sentences in the context of security information. Our algorithm's efficacious performance will be fruitful for the security analysts who were our intended end-users. With the help of our research, we can accentuate the sentences in the APT reports. Our end users can quickly skim through large reports and improve their enterprises' evasion and prevention strategies in times of adversaries.

We believe our research can take a new direction if we improve more on the quality of the dataset. We worked on the available datasets but in the future, we plan on working on a better dataset. We plan to make use of the large model of BERT which has more number of attention layers in it and thus is expected to perform better than the base model used in our approach. We also plan to explore other SubTasks, which revolve on entity extraction and linking with the use of these transformers and make an end-to-end system.